# AN OPTIMUM TIME QUANTUM USING LINGUISTIC SYNTHESIS FOR ROUND ROBIN CPU SCHEDULING ALGORITHM

Supriya Raheja[1], Reena Dhadich[2] and Smita Rajpal[1]

[1]Department of Computer Science & Engineering, ITM University, Gurgaon, India
`Supriya.raheja@gmail.com, smita_rajpal@yahoo.co.in`

[2] Department of MCA, Government Engg. College, Ajmer, Rajasthan, India
`Reena.dadhich@gmail.com`

## ABSTRACT

*In Round Robin CPU scheduling algorithm the main concern is with the size of time quantum and the increased waiting and turnaround time. Decision for these is usually based on parameters which are assumed to be precise. However, in many cases the values of these parameters are vague and imprecise. The performance of fuzzy logic depends upon the ability to deal with Linguistic variables. With this intent, this paper attempts to generate an Optimal Time Quantum dynamically based on the parameters which are treated as Linguistic variables.  This paper also includes Mamdani Fuzzy Inference System using Trapezoidal membership function, results in LRRTQ Fuzzy Inference System. In this paper, we present an algorithm to improve the performance of round robin scheduling algorithm. Numerical analysis based on LRRTQ results on proposed algorithm show the improvement in the performance of the system by reducing unnecessary context switches and also by providing reasonable turnaround time.*

## 1. INTRODUCTION

In multitasking and multiprogramming environment the way of dispatching the processes to the CPUs is called process scheduling. The main goal of the scheduling is to maximize the performance of the system as well as to minimize response time, waiting time, turnaround time and also the number of context switches. When there are number of processes in the ready queue, the algorithm which decides the order of execution of the processes is called a scheduling algorithm. There are various CPU scheduling algorithms have been defined such as First Come First Served FCFS, Shortest Job First (SJF), Shortest Remaining Time Next (SRTN), Round Robin (RR) scheduling algorithm. In this work our concern is with Round Robin scheduling algorithm. RR is designed especially for time-sharing systems. In RR every process has equal priority and is given a fixed time quantum. Every process got CPU only for this time quantum after which the process is preempted. It provides improved response time as compare to other scheduling algorithms. But increased waiting time and turnaround time increased due to use of constant time quantum.  All the decisions for the size of time quantum are usually based on the crisp parameters. Sometimes in many cases these parameters may be vague or imprecise.

The binary logic of modern computers usually cannot describe the vagueness or impreciseness of the real world. Computers do not think as human brains do. A computer "thinks" only the precise facts in binary form and statements in binary logic that are either true or false. With this demarcation, the human brain can think with vague statements that involve uncertainties or imprecise values like: "The more number of processes," or "Burst time is small" or "Time quantum is large." Unlike computers, humans have common sense that makes them to intellect in a world where things are partially true. Prof. Zadeh provides solution to this problem by giving the concept of Fuzzy logic [8].

Linguistic Variables are the primal part of the fuzzy logic. Linguistic variables can be used in every important expect by human beings. A Linguistic variable can be defined as a variable whose values are words rather than numbers.

In this paper, we are generating an Optimal Time Quantum using fuzzy logic. In our work the system adjusts the time quantum according to average burst time & the number of processes present in the ready queue. These two parameters are treated as Linguistic Variable. The resulted time quantum itself represented as Linguistic variable and determined value of time quantum after defuzzification by the system is optimum which improves the larger waiting time & turnaround time for processes in RR Scheduling. We are also proposing an algorithm which further improves the performance of system.

## 2. PRELIMINARIES

This section discusses the basic preliminaries of fuzzy Logic and Fuzzy Inference System.

### 2.1. Fuzzy Logic

Fuzzy logic reflects how human think and behave. Fuzzy logic is a superset of Boolean logic, handles the concept of partial truth i.e. truth values between "completely true" and
"completely false". As its name suggests, it is the logic of reasoning which are approximate rather than exact [6, 8]. The importance of fuzzy logic lies on the fact that most modes of human reasoning are approximate in nature. In standard set theory, an object either does or does not belong to a set. The number three belongs fully to the set of odd numbers and not at all to the set of even numbers; can't an object belong to both a set means number three cannot belongs to the set of odd numbers as well as to the set of even numbers. Fuzzy Logic preserves the structure of logic and provides a way that an object can belong to both sets at the same time. In fuzzy sets objects belongs only partially to a fuzzy set. They may also belong to more than one set. If there are total 10 processes in the system then it is uncertain that the system with 5 processes either lie in the "system having less number of processes" or in the "system having more number of processes".  It attempts to model human sense of words and decision making by using the concept of Linguistic Variables.

Linguistic variables are the variables of the fuzzy system whose values are words or sentences from a natural language, instead of numbers used in classical logic.  These variables represent crisp information in a form that is appropriate for the problem. A linguistic variable is generally decomposed into a set of linguistic terms. Like in our work linguistic variable "LNOP" represent the number of processors. To qualify the number of processors terms such as "fewer", "more" are used in real life. These are the linguistic values of "LNOP". To quantify the linguistic term a member function is used.

## 2.2. Fuzzy Set Theory

Fuzzy sets are sets whose elements have degrees of membership. Fuzzy Set Theory was introduced by Professor Lotfi Zadeh in 1965 [8].Over the classical set theory, Zadeh introduced the concept of fuzzy set theory which has been applied almost all the fields such as computer sciences, medical sciences, to solve the mathematical problems, expert systems and many more. In classical set theory, the membership of elements in a set is assessed in binary terms - an element either belongs or does not belong to the set. By contrast, fuzzy set theory permits the gradual assessment of the membership of elements in a set; this is described with the aid of a membership function valued in the real unit interval [0, 1].

Let $X = \{u_1, u_2, ....., u_n\}$ be the universe of discourse. The membership function $\mu_A(u)$ of a fuzzy set A is a function $\mu_A : X \rightarrow [0,1]$. A fuzzy set A in X is defined as the set of ordered pairs $A = \{(u, \mu_A(u)) : u \in X \}$, where $\mu_A(u)$ is the grade of membership of element u in the set A[8]. The greater $\mu_A(u)$, the greater "the element u belongs to the set A" [8]. Prof. Zadeh generalized classical set theory by developing the concept of fuzzy set theory.

## 2.3. Fuzzy Inference System

The key unit of the fuzzy logic system is Fuzzy Inference System. The principal work of this system is decision making. Fuzzy inference is the process of articulating the mapping from a given input to an output using fuzzy logic [9]. This mapping provides a base from which decisions can be made.

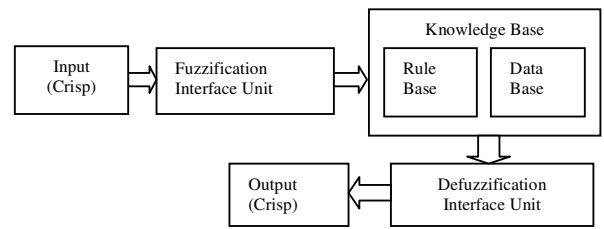

Figure 1. Block Diagram of FIS

The process of fuzzy inference requires all information's regarding membership functions, Logical operations, and If-Then Rules. There are mainly two types of Fuzzy Inference Systems: Mamdani-type and Sugeno-type [9]. Mamdani's fuzzy inference method is the most ordinarily used fuzzy methodology.

In Mamdani-type inference the output membership functions are fuzzy sets. After the aggregation process, there is a fuzzy set for each output variable that further needs defuzzification as shown in above figure 1. Sugeno's inference system uses a single spike as the output membership function rather than a distributed fuzzy set. It increases the efficiency of the defuzzification process. The only difference between these two methods is in the output. The Sugeno's output membership functions are either linear or constant but Mamdani's inference process considers the output membership function to be fuzzy sets [9].

In this work trapezoidal membership function is used. A trapezoidal membership function is described by a Quadruple A (a, b, c, d) as shown in figure 2.

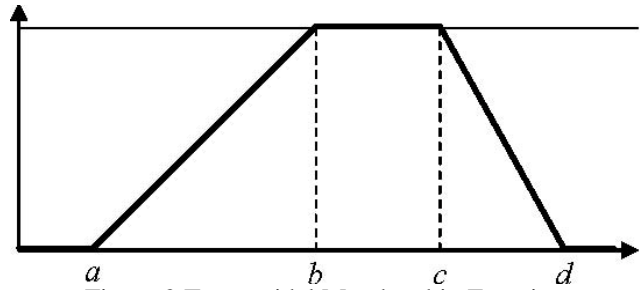
Figure 2 Trapezoidal Membership Function

In general, a trapezoidal fuzzy number is defined by the following membership function:

$$\mu(x) = \begin{cases} 1 - \dfrac{a-x}{c} & \text{if } a-c \le x < a \\ 1 & \text{if } a \le x \le b \\ 1 - \dfrac{x-b}{d} & \text{if } b < x \le b+d \\ 0 & \text{otherwise} \end{cases}$$

## 3. RELATED WORK

Since RR is used in almost every operating system. Many researchers have done worked on this. Based on the linguistic synthesis very little work on scheduling algorithms is available. The works described here is not defined in the form of linguistic synthesis.

The static time quantum which is a limitation of RR was removed by taking dynamic time quantum by Matarneh[1]. Matarneh [1] founded that a dynamic time quantum could be calculated by the median of burst times for the set of processes in ready queue. In such case, the quantum value must be modified to avoid the overhead of context switch time [1].

In this paper, author proposed a new algorithm AN, the idea of this approach is to make the operating systems adjusts the time quantum according to the burst time of the set of waiting processes in the ready queue.

Samih M. Mostafa [2] proposed a method using integer programming to solve equations that decide a value that is neither too large nor too small such that every process has reasonable response time and the throughput of the system is not decreased due to un-necessarily context.

## 4. PROPOSED FIS FOR OPTIMUM TIME QUANTUM (LRRTQ)

In this paper we are generating the Linguistic Optimum Time Quantum (LOTmQm) by designing Mamdani Fuzzy Inference System with trapezoidal membership function named LRRTQ using MATLAB tool. In our designed FIS, range of possible values for the input and output variables are determined.

We are using two Linguistic variables for inputs, "LNOP" that holds the total number of processes available in ready queue and "LABT" that gives the information of average burst time of all the processes available in the ready queue. As we are using Mamdani system the output is further a fuzzy set.  Our designed FIS generates one output i.e. Optimum Time Quantum and is represented by Linguistic Variable "LOTmQm" as shown below in figure 3.

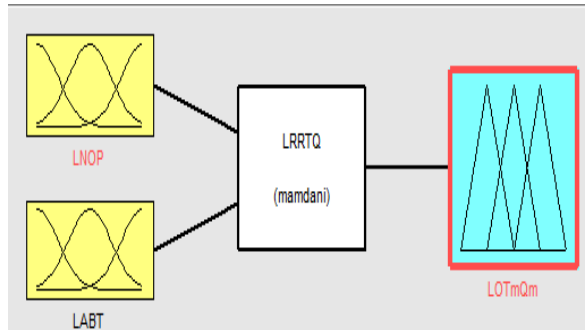

Figure 3. Proposed FIS – LRRTQ

A Linguistic variable is decomposed into linguistic terms. In LRRTQ, the linguistic values for the defined linguistic variables are shown in Table 1 like Linguistic Variable "LABT" having three values "small", "Average" and "large".  These linguistic values are quantified using member functions. In our work we have assigned three membership functions (MF1, MF2, MF3) to each of the input variable as well as to the output variable described in Table 1.

Table 1. Membership Functions

|  | **LNOP** | **LABT** | **LOTmQm** |
|---|---|---|---|
| **Type** | Trapezoidal | Trapezoidal | Trapezoidal |
| **Range** | 1-10 | 1-12 | 1-5 |
| **MF1** | Fewer<br>[-2 0.5 1.5 4] | Small<br>[-4 0.4 1.5 4] | Small<br>[0 0.7 1.4 2.1] |
| **MF2** | Ordinary<br>[3  4.8  5.5  7.5] | Average<br>[3 5.5 6.5 9] | Medium<br>[1.5 2.2 2.8 3.8] |
| **MF3** | More<br>[7 9 10.5 12] | Large<br>[7.5 10 11 13.5] | Large<br>[3.5 4.5 5 6] |

Additionally   we are defining member functions graphically as shown in figure 4(a) and figure 4(b) for input linguistic variables [LNOP, LABT] & figure 5 for output variable LOTmQm respectively.

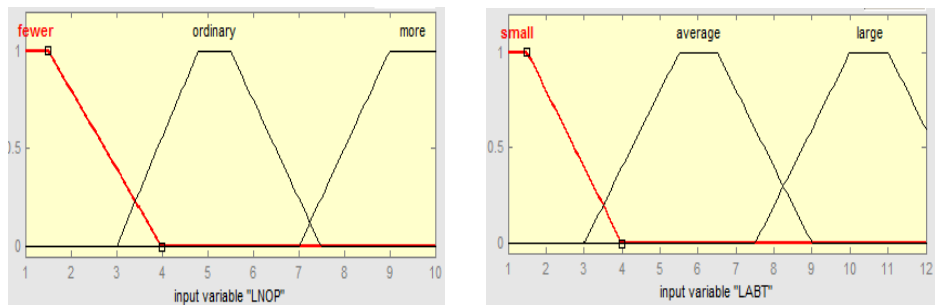

Figure 4(a)-4(b) Membership functions for linguistic variable LNOP and LABT

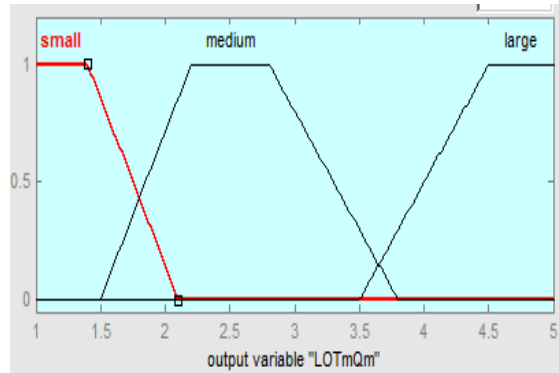

Figure 5. Membership function for Linguistic variable LOTmQm

## 4.1. Rule Base for LRRTQ

Rules are formed keeping in mind intuitive relationship between input and output parameters. In the Fuzzy Inference System, rule base contains the fuzzy rules that provide us the way to combine the input parameters to the output parameters as they are connected in real worlds. In our Rule Base, variable LNOP having 3 fuzzy ranges fewer, ordinary and more, variable LABT also having 3 fuzzy ranges small, average, and large that gives a rule base matrix with size 3x3= 9 "if-then" rules as shown in figure 6. Based on these rules the system LRRTQ will generates the Optimum Time Quantum i.e. the output LOTmQm.

```
1. If (LNOP is fewer) and (LABT is small) then (LOTmQm is small) (1)
2. If (LNOP is fewer) and (LABT is average) then (LOTmQm is medium) (1)
3. If (LNOP is fewer) and (LABT is large) then (LOTmQm is large) (1)
4. If (LNOP is ordinary) and (LABT is small) then (LOTmQm is small) (1)
5. If (LNOP is ordinary) and (LABT is average) then (LOTmQm is medium) (1)
6. If (LNOP is ordinary) and (LABT is large) then (LOTmQm is medium) (1)
7. If (LNOP is more) and (LABT is small) then (LOTmQm is small) (1)
8. If (LNOP is more) and (LABT is average) then (LOTmQm is small) (1)
9. If (LNOP is more) and (LABT is large) then (LOTmQm is medium) (1)
```

Figure 6. Rule base matrix for LRRTQ

## 4.2 Implementation Results

Through rule viewer tool we've reached to a conclusion that our designed system using fuzzy logic could generate the Optimum Time Quantum for Round Robin by changing the idea of fixed time quantum to dynamic based on Linguistic synthesis i.e. using two linguistic variables "LNOP" and "LABT" automatically without the interfere of user. When the Number of Processors "LNOP" is 4 and the Average Burst Time "LABT" is 6 then the generated Optimum Time Quantum "LOTmQm" by LRRTQ is 2.61. As the LNOP changed to 3 and the LABT to 8 the value of LOTmQm is 3.06 for new inputs. This can be seen through the surface view as shown in figure 7. Here we could say that our system has generated dynamically different values of time quantum based on different inputs.

International Journal on Soft Computing ( IJSC ) Vol.3, No.1, February 2012

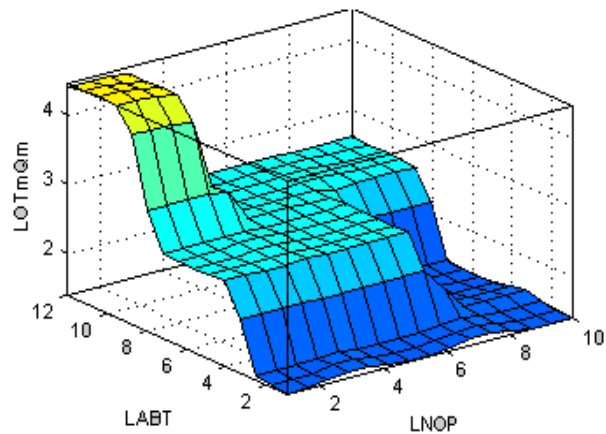

Figure 7. Surface View for LRRTQ

## 5. PROPOSED ALGORITHM

In our proposed algorithm, the processes are sorted in increasing order of their burst time so that shortest process will remove earlier from the ready queue to give better turnaround time and waiting time. Whenever a process comes, the required burst time is compared with the available processes in the ready queue and accordingly ready queue is updated. The performance of RR algorithm entirely depends on the size of time quantum. In our proposed algorithm time quantum is taken from our designed FIS, LRRTQ. This dynamic time quantum is used by all processes present in ready queue and this continues up to complete execution until a new process arrives in a ready queue. In succeeding cycles if new process arrives then time quantum is again calculated taking into consideration of all the parameters. In our algorithm, the processes which are already present in the ready queue, their arrival time considered as zero.

### 5.1. Assumptions

We assumed that the entire processes are non cooperative processes. Attributes like burst time, number of processes are known before submitting the processes. Time quantum should be larger than the total swap time.  Different terminologies such as LNOP, LABT and LOTmQm are used in the proposed algorithm that represents number of processes available in ready queue, Abstract Burst Time and Time Quantum respectively.

### 5.2 Algorithm

Step1: Sort the ready queue in increasing order according to their burst time.
Step2: Count the LNOP.
Step3: While (ready queue!=NULL)
Step4: Calculate the average burst time LABT for all processes available in ready queue.
Step5: Find the LTQ using FIS –LRRTQ.
Step6: Apply Round Robin algorithm using LOTmQm until new process arrived in ready queue.
Step7: If any new process arrive in ready queue.
Step8: Repeat all the above steps & update the LNOP & LABT.
Step9: Finally calculate the waiting time, turnaround time and context switches

## 6. NUMERICAL ANALYSIS

We have taken different cases for evaluating our work as discussed below:

**Case 1:** Assume four processes arrived at time unit 0 with burst time (P1 = 8, P2 = 5, P3 = 4, P4 = 7):

Table 2. Case 1

| Input [LNOP,LABT] [4, 6] | Output[$LOT_mQ_m$] 2.6 | |
|---|---|---|
| | RR algorithm | Proposed Algorithm |
| Average Waiting Time | 15.3 | 11.8 |
| Average Turnaround Time | 21.3 | 17.8 |
| Context Switch | 10 | 10 |

**Case 2:** Assume three processes arrived at time unit 0 with burst time (P1 = 8, P2 = 10, P3 =6):

Table 3. Case 2

| Input [LNOP,LABT] [3, 8] | Output Fixed Time Quantum 2.6 | | Output [$LOT_mQ_m$] 3.0 | |
|---|---|---|---|---|
| | RR | Proposed | RR | Proposed |
| Average Waiting Time | 14.5 | 11.8 | 12.8 | 10.7 |
| Average Turnaround Time | 22.5 | 19.8 | 17.3 | 18 |
| Context Switch | 9 | 9 | 8 | 8 |

From the above numerical analysis it is clear that the Optimum Time Quantum approach is more effective rather than the fixed time quantum and by applying Optimum Time Quantum with the proposed algorithm it significantly reduces the context switch, turnaround time and the waiting time. As when the

## 7. CONCLUSIONS

In this paper we have generated an Optimum Time Quantum using Linguistic synthesis for Round Robin scheduling which improves the performance of system as compare to system using fixed time quantum. We have used the MATLAB tool to design the LRRTQ Fuzzy Inference System using Trapezoidal membership functions to determine the Optimum Time Quantum, LOTmQm. Additionally, we have proposed an algorithm which further enhances the performance of system. Numerical analyses based on implementation results conclude that it significantly reduces the context switch, turnaround time and the waiting time proposed on different cases in section 6.

## ACKNOWLEDGEMENTS

Author would like to thank Dr. Reena Dadhich who provided invaluable comments in. Author would like to thank Dr. Smita Rajpal for her inputs in the conceptualization of the paper and her support throughout for technical discussions.

## Authors

**Supriya Raheja**, ITM University, pursuing her PhD in Computer Science from Banasthali University. She is specialized in OOPs, Operating System and Networks. She is working as a Reviewer/Committee member of various International Journals and Conferences. Her total Research publications are 11. She has attended & organized various faculty development programs and workshops.

**Dr. Reena Dadhich** is presently working as an Associate Professor and Head of the Department of Master of Computer Applications at Engineering College Ajmer, India. She received her Ph.D. (Computer Sc.) and M.Sc. (Computer Sc.) degree from Banasthali University, India. Her research interests are Algorithm Analysis & Design ,Wireless Ad-Hoc Networks and Software Testing. She has more than 12 years of teaching experience. She is working as an Editorial Board Member / Reviewer/Committee member of various International Journals and Conferences. She has written many research papers and books.

**Dr. Smita Rajpal**, ITM University, completed her PhD in Computer Engineering. She has a total work experience of 11 years. She is specialized in TOC, Compiler Design, Soft Computing and RDBMS. She is a Java certified professional. She is working as an Editorial Board Member / Reviewer/Committee member of various International Journals and Conferences. She is an active member of IEEE. Her biography is a part of Marquis who's who in the world, 2010. Her total Research publications are 22 and book chapter's-5.She has published three books.